\documentclass[preprint,12pt]{elsarticle}

\usepackage{amssymb}

\usepackage[version=3]{mhchem} 
\usepackage{graphicx}
\usepackage{float} 
\usepackage{textcomp,mathcomp} 
\usepackage{siunitx}
\usepackage{amsmath} 
\usepackage{arydshln} 

\journal{Applied Surface Science}

\begin{document}

\begin{frontmatter}



\title{Laser-Synthesized Amorphous PdSe\textsubscript{2-x} Nanoparticles: A Defect-Rich Platform for High-Efficiency SERS, Photocatalysis, and Photothermal Conversion}


\author[label1]{Andrei Ushkov}
\author[label1,label2]{Nadezhda Belozerova}
\author[label1]{Dmitriy Dyubo}
\author[label1]{Ilya Martynov}
\author[label1,label3,label4]{Alexander Syuy}
\author[label1,label5]{Daniil Tselikov}
\author[label3]{Georgy Ermolaev}
\author[label1]{Sergey V. Bazhenov}
\author[label6]{Roman I. Romanov}
\author[label3]{Ivan Kruglov}
\author[label5]{Anton A. Popov}
\author[label1]{Alexander Chernov}
\author[label1,label7,label8,label9]{Alexey D. Bolshakov}
\author[label1]{Sergey Novikov}
\author[label1,label3]{Andrey A. Vyshnevyy}
\author[label1,label3]{Aleksey Arsenin}
\author[label10]{Andrei V. Kabashin}
\author[label3]{Gleb I. Tselikov}
\author[label3]{Valentyn Volkov}

\affiliation[label1]{organization={Moscow Center for Advanced Studies},
            addressline={Kulakova Str. 20},
            city={Moscow},
            postcode={123592},
            country={Russia}}
\affiliation[label2]{organization={Frank Laboratory of Neutron Physics, Joint Institute for Nuclear Research},
            addressline={Joliot-Curie 6},
            city={Dubna},
            postcode={141980},
            country={Russia}}
\affiliation[label3]{organization={Emerging Technologies Research Center, XPANCEO, Internet City, Emmay Tower},
            addressline={Al Sufouh 2}, 
            city={Dubai},
            postcode={123592}, 
            country={United Arab Emirates}}
\affiliation[label4]{organization={Department of General Physics, Perm National Research Polytechnic University},
            city={Perm},
            postcode={614990},
            country={Russia}
}           
\affiliation[label5]{organization={MEPhI, Institute of Engineering Physics for Biomedicine (PhysBio)},
            addressline={Kashirskoe shosse 31}, 
            city={Moscow},
            postcode={115409}, 
            country={Russia}}
\affiliation[label6]{organization={MEPhI, Department of Solid-State Physics and Nanosystems},
            addressline={Kashirskoe shosse 31}, 
            city={Moscow},
            postcode={115409}, 
            country={Russia}}           
\affiliation[label7]{organization={Faculty of Physics, St. Petersburg State University},
            addressline={Universitetskaya Emb. 7--9}, 
            city={St. Petersburg},
            postcode={199034}, 
            country={Russia}}
\affiliation[label8]{organization={Alferov University, Russia},
            addressline={Khlopina 8/3}, 
            city={St. Petersburg},
            postcode={194021}, 
            country={Russia}}
\affiliation[label9]{organization={Laboratory of Advanced Functional Materials, Yerevan State University},
            addressline={1 Alek Manukyan St.}, 
            city={Yerevan},
            postcode={0025}, 
            country={Armenia}}
\affiliation[label10]{organization={Aix-Marseille University, CNRS, LP3},
            addressline={Av. de Luminy 163}, 
            city={Marseille},
            postcode={13288}, 
            country={France}}
\begin{abstract}
The control of material properties at the atomic scale remains a central challenge in materials science. Transition metal dichalcogenides (TMDCs) offer remarkable electronic and optical properties, but their functionality is largely dictated by their stable crystalline phases. Here we demonstrate a single-step, ligand-free strategy using femtosecond laser ablation in liquid to transform crystalline, stoichiometric palladium diselenide (PdSe\textsubscript{2}) into highly stable, amorphous, and non-stoichiometric nanoparticles (PdSe\textsubscript{2-x}, with x$\approx$1). This laser-driven amorphization creates a high density of selenium vacancies and coordinatively unsaturated sites, which unlock a range of emergent functions absent in the crystalline precursor, including plasmon-free surface-enhanced Raman scattering with an enhancement factor exceeding 10$^6$, a 50-fold increase in photocatalytic activity, and near-infrared photothermal conversion efficiency reaching 83\%. Our findings establish laser-induced amorphization as a powerful top-down approach for defect-engineered TMDCs and advances their practical usage in optics, catalysis, and nanomedicine.
\end{abstract}




\end{frontmatter}



\section{Introduction}
The "materials by design" paradigm largely relies on effective synthesis routes and the ability to manipulate the atomic structure to unlock novel or enhanced functionalities for energy, photonics, and medicine. Two-dimensional (2D) van der Waals (vdW) materials \cite{duong2017van,gogotsi2023rise}, particularly transition metal dichalcogenides (TMDCs), became an outstanding platform for the concept realization, offering highly tunable electronic \cite{hossen2024defects,falin2021mechanical}, optical \cite{sun2015electronic}, and mechanical \cite{qi2023recent,huang2023general} properties.

Recently, platinum group metal chalcogenides, e.g., Pd, Pt, and Rh, have attracted attention as promising materials in catalysis \cite{saraf2018pursuit}, electronic semiconducting devices \cite{zeng2020van} and thin film absorbers for solar panels \cite{alfieri2023ultrathin}. A distinctive feature of platinum group metals is the filled or nearly filled d orbital, leading to the (semi)metallic behavior of their chalcogenides and a high electrical conductivity.

Among these, palladium diselenide (PdSe\textsubscript{2}) stands out due to its unusual puckered pentagonal lattice structure, in-plane optical anisotropy \cite{gu2020two} and high air stability \cite{oyedele2017pdse2} making it a promising candidate for near-infrared optoelectronics \cite{liang2019high,zeng2019controlled,wang2021applications} and catalysis. Interestingly, most of the Pd-Se studies are concentrated in electrocatalysis \cite{hu2023synthesis,kukunuri2017effects,lin2021planar,qin2022extraordinary}, whereas only a few are devoted to SERS applications \cite{lei2022enhanced,jena2023evidence,zhang2023general}. The reason for it are quite moderate enhancement factors (EF) due to the chemical (non-plasmonic) SERS mechanism of the crystalline PdSe\textsubscript{2}, far below the EF$>10^6$ routinely obtained with plasmonic nanostructures. An interesting approach for SERS improvement is a defect engineering \cite{jena2023evidence} in bilayer PdSe\textsubscript{2} nanosheets by keeping the overall structure, however, predominantly crystalline. In this context, the material properties are still largely dictated by its thermodynamically stable, ordered crystalline forms, including PdSe\textsubscript{2} \cite{ibrahim2022phase}, Pd\textsubscript{7}Se\textsubscript{2} \cite{hu2023synthesis}, Pd\textsubscript{4}Se \cite{singh2012palladium}, Pd\textsubscript{17}Se\textsubscript{15} \cite{ibrahim2022phase}.

This crystalline constraint inspires a question: what new properties could be accessed in the amorphous counterpart of the material? The abundance of Pd-Se systems indicated above suggests the existence of a stable amorphous phase of palladium selenide with inherently high density of defects and coordinatively unsaturated sites. Such a material should possess excellent SERS and catalytic performance at the coordinatively unsaturated surface. The primary obstacle, however, is the synthesis and stabilization of these amorphous TMDC phases, because conventional bottom-up (CVD) and top-down (liquid-phase exfoliation) approaches are optimized for crystallinity. Thus, a method capable of driving a material far from its thermodynamic equilibrium is required to access and stabilize these metastable amorphous states.

Here, we introduce femtosecond pulsed laser ablation in liquid (PLAL) as a unique material processing tool to overcome this challenge. Using PLAL we show that an ordered, stoichiometric crystalline PdSe\textsubscript{2} target can be controllably converted into a stable, disordered, non-stoichiometric, and highly functional amorphous nanomaterial, PdSe\textsubscript{2-x}. The obtained nanoparticles offer significant advantages over conventional planar plasmon-free substrates due to the synergetic combination of mechanisms: a large coordinatively unsaturated surface, defect-state energy levels for improved charge transfer and plasmon-free hotspots in nanoparticle aggregates. In contrast to the reported amorphization in "wet" chemistry, we use a non-demanding, simple one-step method that does not require a glovebox with an inert atmosphere \cite{wu2025synthesis} or high temperatures \cite{yu2023low} with lengthy stirring. In addition, the proposed method allows direct manufacturing scaling via a continuous-flow reactor \cite{wu2023real}.




\section{Results and discussion}
\subsection{Structural analysis of palladium selenide nanoparticles and flakes}
\label{struct_analysis}

\begin{figure}[H]
\centering\includegraphics[width =1\textwidth]{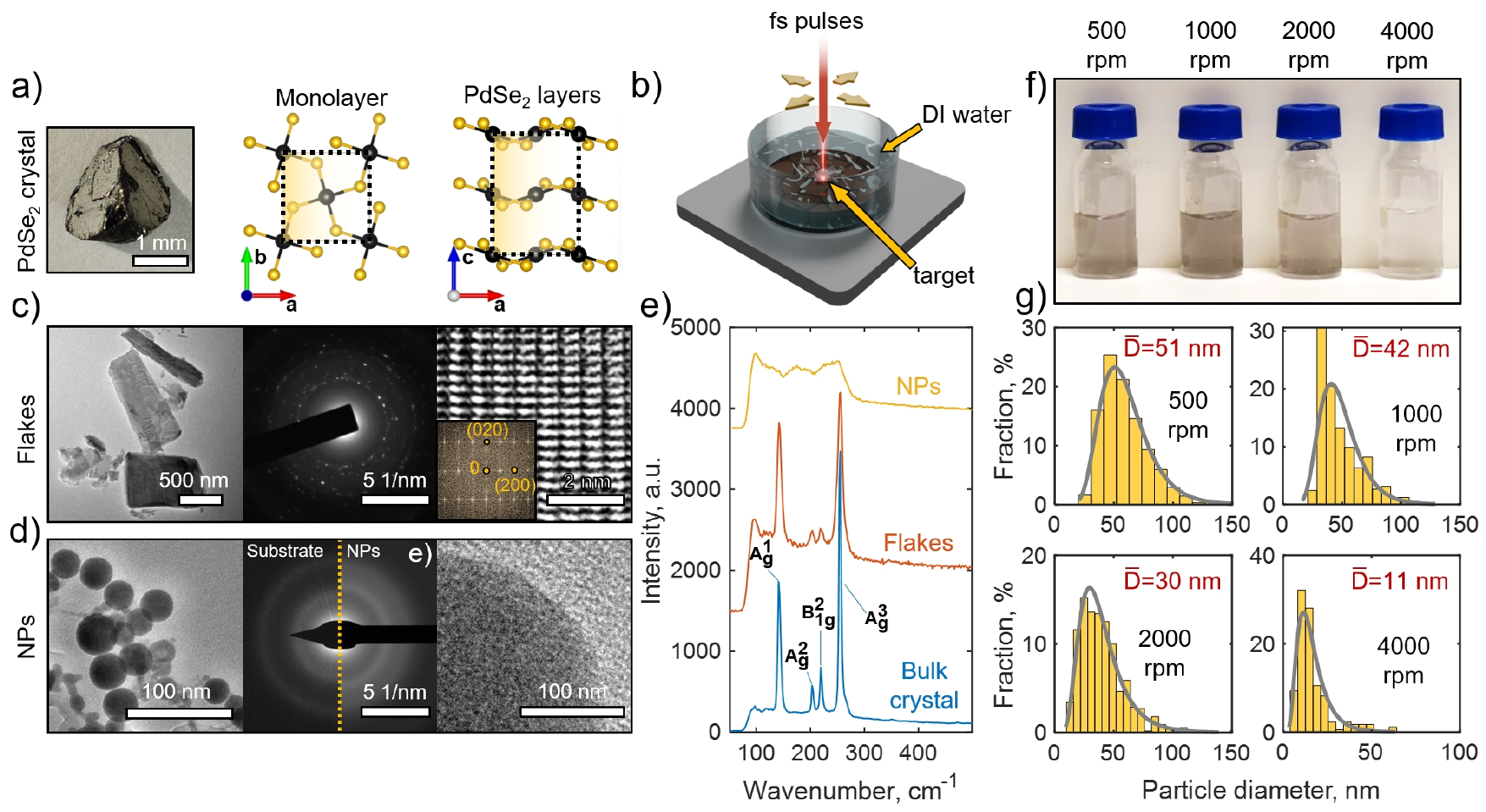}
\caption{(a) Photograph of PdSe\textsubscript{2} crystal used as a target and its crystal structure, the unit cell is denoted as a dashed box; (b) Schematic view of PLAL; (c) From left to right: typical TEM image and SAED of PdSe\textsubscript{2} flakes, and a high-resolution TEM photograph of a zoomed area of a single flake proving its crystalline structure; (d) From left to right: typical TEM image and SAED of PdSe\textsubscript{2-x} nanoparticles, and a zoomed TEM image of a single nanoparticle edge; (e) Raman spectra of bulk crystal, crystalline flakes and amorphous nanoparticles; (f) Photographs of colloids separated by centrifugation at different rotation speeds; (g) Size distributions and average sizes of PdSe\textsubscript{2-x} nanoparticles, obtained at different rotational speeds and measured by counting on TEM image.}
\label{Article1}
\end{figure}

Figure \ref{Article1} outlines the structural properties of nanoparticles, synthesized via femtosecond laser ablation in DI water (refer to the Methods section for detailed information). A high-purity (>99.999\%) commercial PdSe\textsubscript{2} crystal (SixCarbon Technology Co., Limited, Shenzhen) is used as a target and a single precursor for pulsed laser nanoparticle synthesis, see Figs.~\ref{Article1}a,b. Micron-sized flakes (Fig.~\ref{Article1}c), produced via the ultrasound-assisted liquid phase exfoliation method \cite{chavalekvirat2024liquid}, preserve the crystallinity of the bulk source, which is confirmed by diffraction rings in SAED and high-resolution TEM images in Fig.~\ref{Article1}c. The measured lattice constants $a=5.76$ \r{A}, $b=5.90$ \r{A} agree well with the experimental values and DFT calculations ($a=5.75$ \r{A}, $b=5.87$ \r{A}) \cite{sun2015electronic,oyedele2017pdse2,soulard2004experimental}. In contrast, the nanoparticles produced via pulsed laser ablation demonstrate an amorphous structure (Fig.~\ref{Article1}d): a single-particle SAED demonstrates amorphous halos, different from those of TEM amorphous carbon film. In addition, the nanoparticles have abrupt edges without an onion-like shell typically found in crystalline TMDC NPs \cite{tselikov2022transition}.

Raman spectroscopy (Fig.~\ref{Article1}e) provides further evidence of the NPs amorphous state, in contrast to both bulk PdSe\textsubscript{2} crystal and flakes. The Raman spectra of the latter samples exhibit all characteristic vibrational modes of PdSe\textsubscript{2} \cite{jena2024salt,jena2023evidence,abdul2024resonance}, as shown in Figure \ref{Article1}e. The spectra have strong, well-defined peaks at 142 cm\textsuperscript{-1} and 256 cm\textsuperscript{-1}, which are assigned to the $A^{1}$\textsubscript{g} and $A^{3}$\textsubscript{g} Raman-active modes, respectively. Furthermore, we observe two distinct modes at 203 cm\textsuperscript{-1} ($A^{2}$\textsubscript{g}) and 220 cm$^{-1}$ ($B^{2}$\textsubscript{1g}), consistent with previous reports. The identical Raman features observed in both the initial crystal and derived flakes confirm their common crystalline PdSe\textsubscript{2} structure, ruling out amorphous phase formation during the flake preparation process.

The Raman spectrum of PdSe\textsubscript{2-x} NPs exhibits broad overlapping bands that contrast sharply with the well-defined peaks of distinct vibrational modes observed in crystalline PdSe\textsubscript{2} samples (Fig.~\ref{Article1}e). The absence of discrete Raman modes in nanoparticles reflects structural disorder: broken translational symmetry broadens vibrational bands, while non-stoichiometric bonding (Pd:Se about 1:1) suppresses zone-center phonons detectable in crystalline PdSe\textsubscript{2}.

For the photothermal conversion tests (see Section \ref{PCE}), several size fractions of the originally synthesized colloid were obtained via the differential centrifugation method \cite{livshits2015isolation}, see Section \ref{label_materials_and_methods} for details. Interestingly, the darkest solution was obtained after 1000 rpm (Fig.~\ref{Article1}f), suggesting the highest concentration of NPs with a mean size $\sim$40 nm in the synthesized colloid. We obtained mean NPs sizes from 51 nm down to 11 nm at different rotation speeds, as shown in Fig.~\ref{Article1}g.

\begin{figure}[H]
\centering\includegraphics[width =0.8\textwidth]{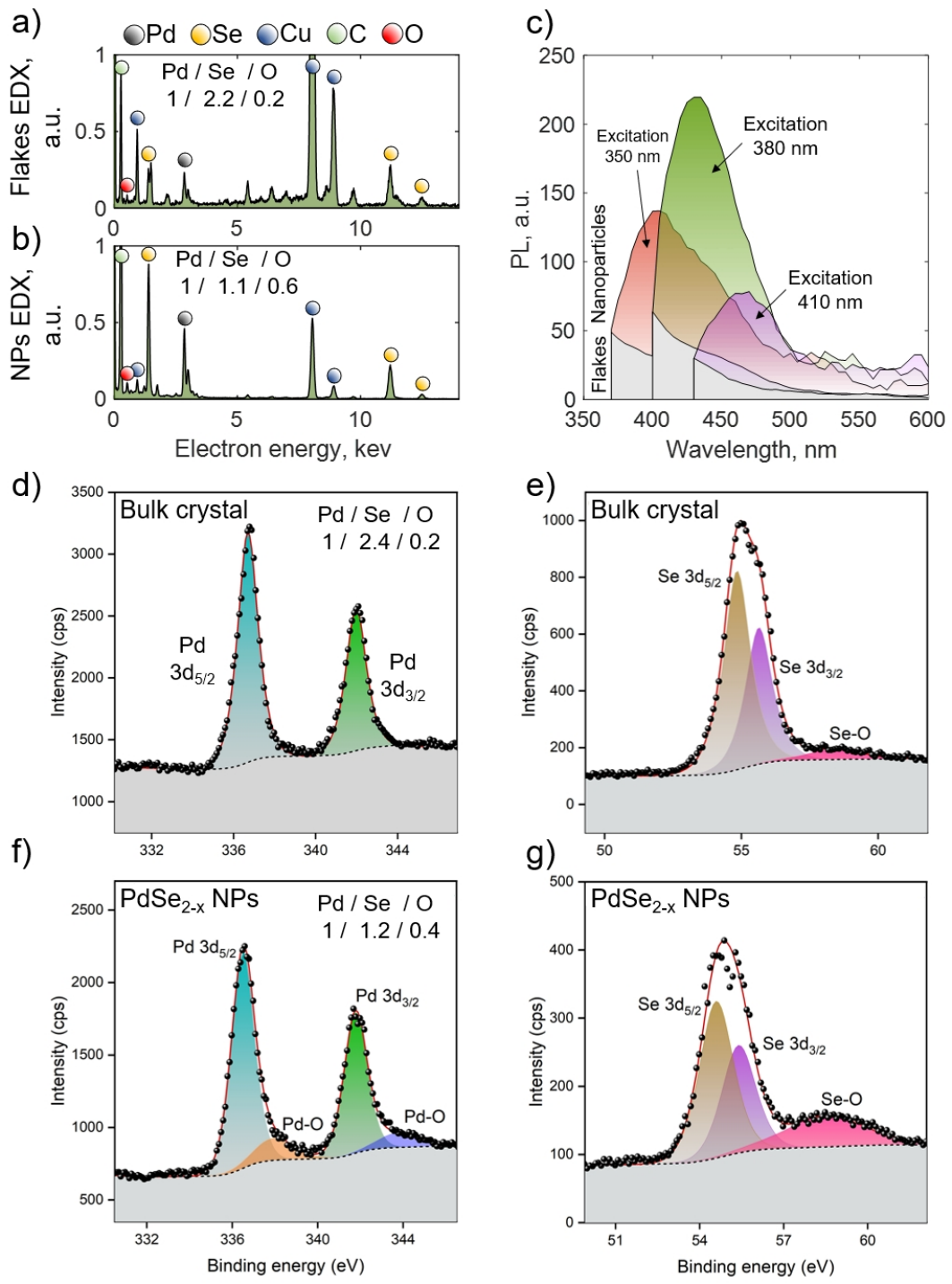}
\caption{ a),b) EDX of (a) flakes of PdSe\textsubscript{2} and (b) nanoparticles of PdSe\textsubscript{2
-x} with TEM photographs in the insets; c) Photoluminescence spectra of PdSe\textsubscript{2} flakes (grey region) and PdSe\textsubscript{2-x} nanoparticles (red, green and blue regions) acquired at excitation wavelengths of 350 nm, 380 nm and 410 nm; d),e) XPS characterization of bulk crystal PdSe\textsubscript{2} for (d) Pd 3d and (e) Se 3d, with shirley baseline fitted spectra; f), g) XPS characterization of PLAL-synthesized PdSe\textsubscript{2-x} NPs for (f)  Pd 3d and (g) Se 3d, with Shirley baseline fitted spectra.}
\label{Article2}
\end{figure}

The synthesized nanoparticles PdSe\textsubscript{2-x} are expected to have pronounced catalytic and Raman enhancement properties in light of recent studies of amorphization-induced electronic transport \cite{zheng2017semiconductor,cong2015noble,chepkasov2025adsorption,fu2023unraveling}. In this regard, stoichiometric analysis is required to understand the type and concentration of active sites/vacancies. The EDX spectra of the flakes, prepared via ultrasound-assisted exfoliation, yield an approximate 1:2 Pd-to-Se ratio for Pd:Se with a minor content of O, see Fig.~\ref{Article2}a. A slight oxidation likely appears during the storage of the prepared flakes in DI water. In contrast, amorphous nanoparticles exhibit noticeably stronger oxidation and have almost equal amounts of Pd and Se atoms (see Fig.~\ref{Article2}b), which resembles Pd\textsubscript{17}Se\textsubscript{15} NPs \cite{ibrahim2022phase}. A significant fraction of incorporated oxygen leads to a high concentration of defects in the nanoparticles, compared to the crystalline PdSe\textsubscript{2} flakes. The presence of defects is also revealed via photoluminescence (PL) measurements. As shown in Fig.~\ref{Article2}c, crystalline flakes have PL spectra with no peaks below the wavelength of 600 nm under excitation at wavelengths of 350-410 nm. It agrees with the minimal energy of 0.9 eV for direct optical transitions in bulk crystalline PdSe\textsubscript{2} \cite{volkov2025all}, which increases up to 1.1-1.3 eV for monolayer flakes \cite{wei2022layer,jena2023evidence}. It leads to the infrared interband recombination and PL peak at $\sim$950 nm \cite{jena2023evidence}. In contrast, the amorphous PdSe\textsubscript{2-x} NPs have a slight PL signal in the visible range (Fig.~\ref{Article2}c). The signal maximum shifts towards larger wavelengths together with excitation wavelengths because of the excitation of various defect states inside the nanoparticles.

For catalytic applications and Raman scattering enhancement, surface defects play a major role. In this regard, in addition to the EDX analysis of particle volume, we performed XPS surface analysis, see Figs.~\ref{Article2}d-g. For crystalline PdSe\textsubscript{2}, the binding energies of Pd 3d\textsubscript{3/2} (341.97 eV), Pd 3d\textsubscript{5/2} (336.69 eV PdO\textsubscript{x}/Pd), and binding energies of Se 3d\textsubscript{3/2} (54.84 eV) and Se 3d\textsubscript{5/2} (55.64 eV) are consistent with the reported values \cite{hoffman2019exploring}. In amorphous PdSe\textsubscript{2-x} NPs the core-level peaks of Pd and Se show a uniform shift toward lower binding energies compared to those of the pristine PdSe\textsubscript{2}; this effect is attributed to the Fermi level lowering upon p-type doping, consistent with previous reports on p-type doping of TMDCs \cite{zhou2024scalable}. Notably, new distinct binding energy peaks associated with Pd 3d core levels at 343.70 eV and 337.90 eV (PdO\textsubscript{2}) and Se 3d core levels at 58.50 eV appear after femtosecond ablation, indicating the formation of Pd-O and Se-O bonds \cite{liang2020performance,jena2023evidence}. The Pd:Se ratio on the surface of nanoparticles is 1:1.2, which is close to the EDX data 1:1.1 (see the insets in Figs.~\ref{Article2}b,f). However, a sufficiently large amount of Se-O (see Supplementary Note 1) indicates the appearance of Se- and O-saturated Se vacancies, preferential for catalytic and Raman-enhanced properties.

\subsection{Photothermal properties of palladium selenide nanoparticles}
\label{PCE}

\begin{figure}[H]
\centering\includegraphics[width =0.9\textwidth]{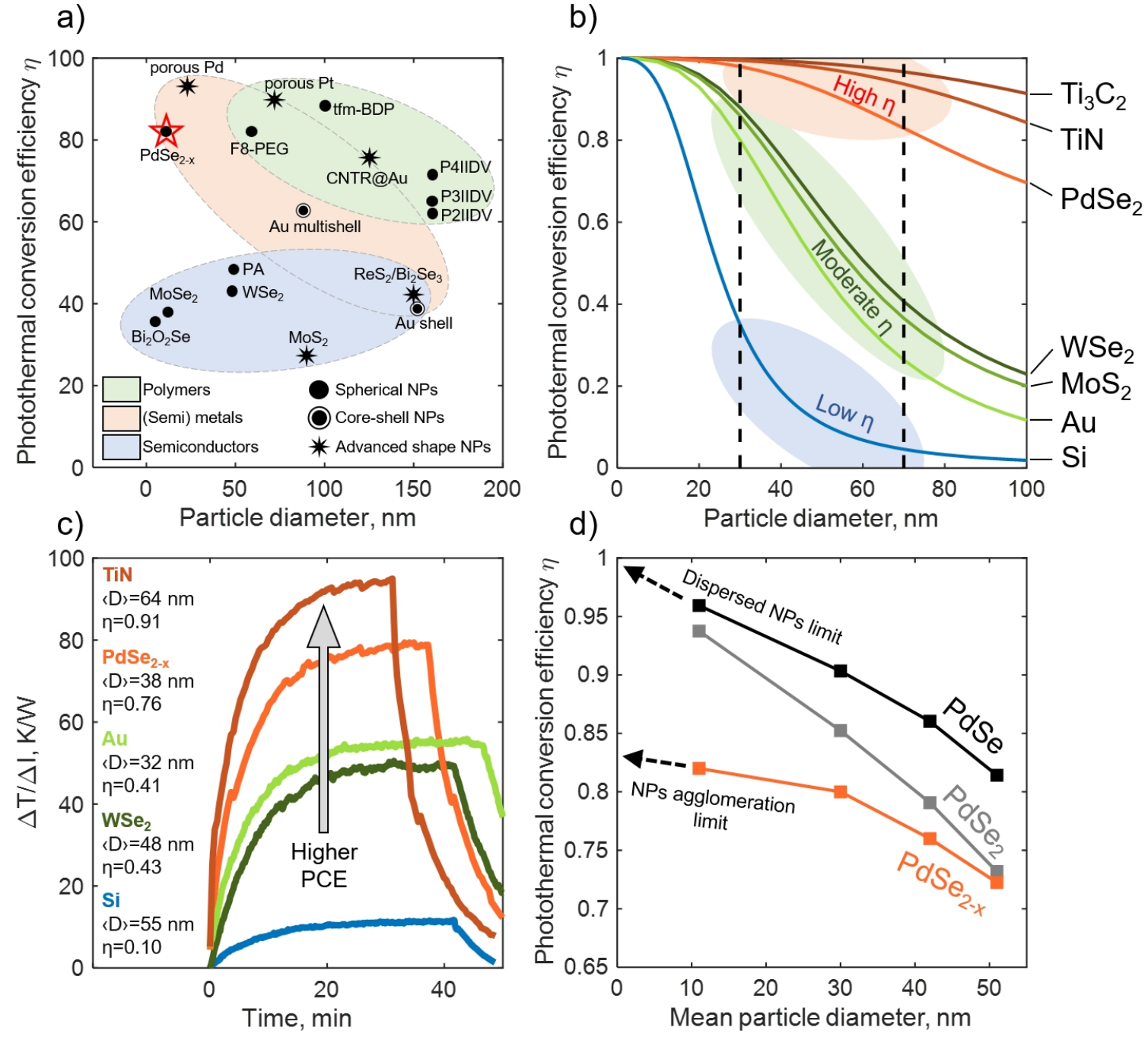}
\caption{Photothermal conversion efficiency (PCE) analysis of nanoparticle systems: (a) Literature survey comparing reported PCE values for various materials versus photothermal agent size; (b) Calculated PCE for sub-100 nm spherical nanoparticles of conventional (Si, Au, TiN) and emerging (MXene Ti\textsubscript{3}C\textsubscript{2}, TMDC MoS\textsubscript{2}) materials at 830 nm (NIR-I window); (c) Experimental PCE values for femtosecond-laser-ablated PdSe\textsubscript{2-x} nanoparticles compared with other materials (TiN, Au, WSe\textsubscript{2}, Si); (d) Theoretical PCE for crystalline PdSe\textsubscript{2} and PdSe nanoparticles versus experimental results for amorphous PdSe\textsubscript{2-x} as a function of particle diameter.}
\label{Article3}
\end{figure}

Photothermal therapy (PTT) is an emerging nanomedicine approach due to the adaptation of novel materials and attempts to perform a combined therapeutic effect. Sub-100 nm NPs are known to be required for their accumulation in tumors due to the enhanced permeability and retention (EPR) effect \cite{ayala2014nanomatryoshkas}. A variety of materials, shapes and hybrid systems have recently been proposed to enhance the photothermal performance of nanoagents in NIR \cite{ushkov2024tungsten,ayala2014nanomatryoshkas,li2019biodegradable,jiang2020polymerization,zhou2013nir,xiao2014porous,xi2020nir,zhu2016porous,song2016gold,ye2023defect,pan2025enhanced,xie2020biodegradable,feng2015flower}. Figure \ref{Article3}a summarizes the experimental findings of these works, paying attention to the measured photothermal conversion efficiency (PCE) and the size of photothermal agents. Metallic and semimetallic systems appear to cover wider ranges of both metrics compared to semiconducting and polymer NPs.
PCE calculations (see Supplementary Note 2) for sub-100 nm spherical nanoparticles made of conventional (Si, Au, TiN) and novel (MXene Ti\textsubscript{3}C\textsubscript{2}T\textsubscript{x}, TMDC MoS\textsubscript{2} and others) materials for NIR-I wavelength 830 nm allow their classification from low- to high PCE, see Fig.~\ref{Article3}b. Among others, crystalline PdSe\textsubscript{2} nanoparticles demonstrate one of the highest PCE in a size range, easily achievable via femtosecond laser ablation. Although the synthesized PdSe\textsubscript{2-x} NPs are amorphous, they retain high photothermal efficiency both in absolute value (measured $\eta=76\%$ for the average NP size of 38 nm) and in comparison with NPs obtained via femtosecond laser ablation of other materials, see Fig.~\ref{Article3}c. All NPs were dispersed in DI water and PCE was measured under 830 nm illumination. Analogously to Fig.~\ref{Article3}b, we performed PCE tests for PdSe\textsubscript{2-x} NPs with different average size fractions, obtained via differential centrifugation (see Section \ref{struct_analysis} for details). Smaller NPs scatter less light, which increases their PCE from 72$\%$ for the average size 51 nm up to 82$\%$ for the average size 11 nm, see Fig.~\ref{Article3}d. In contrast to the theoretically predicted PCE curves for crystalline PdSe\textsubscript{2} and PdSe NPs (see Supplementary Note 3 for refractive index data), the experimental PCE curve for PdSe\textsubscript{2-x} NPs asymptotically tends to the value $\eta\approx0.83<1$ at the zero size limit; it could be caused by NP agglomeration. Nevertheless, the experimentally measured PCE approves the suitability of amorphous PdSe\textsubscript{2-x} for highly efficient photothermal agents in NIR-I transparency window.

\subsection{Palladium Diselenide Flakes and Nanoparticles for SERS}

SERS activity analysis of synthesized PdSe\textsubscript{2-x} nanoparticles and PdSe\textsubscript{2} flakes performed using a series of dyes: rhodamine 6G (Rh6G), rhodamine B (RhB), crystal violet (CV) and methyl orange (MO) with concentrations ranging from $10^{-4}$ to $10^{-10}$ M.
Representative Raman peaks for the entire set of dyes were identified. The Raman spectra of Rh6G molecules obtained from the PdSe\textsubscript{2-x} substrates show prominent peaks at approximately 612, 772, 1127, 1186, 1310, 1360, 1505, 1573, and 1649 cm$^{-1}$ under 532 nm excitation. These peaks align well with previously reported Raman spectra of Rh6G \cite{jebakumari2023engineered,lv2022femtomolar}. 
The SERS spectra of crystal violet obtained with 633 nm excitation clearly show characteristic Raman bands at 425, 916, 802, 1175, 1375, and 1619 cm$^{-1}$, consistent with data from the literature \cite{bonse2020laser}. Under 532 nm excitation, the most intense Raman peaks were observed at 622, 1198, 1279, 1358, 1506, 1529, and 1646 cm$^{-1}$, which are characteristic of RhB \cite{becher2023formation,ushkov2024tungsten,he2012surface}.

\begin{figure}[H]
\centering\includegraphics[width =1\textwidth]{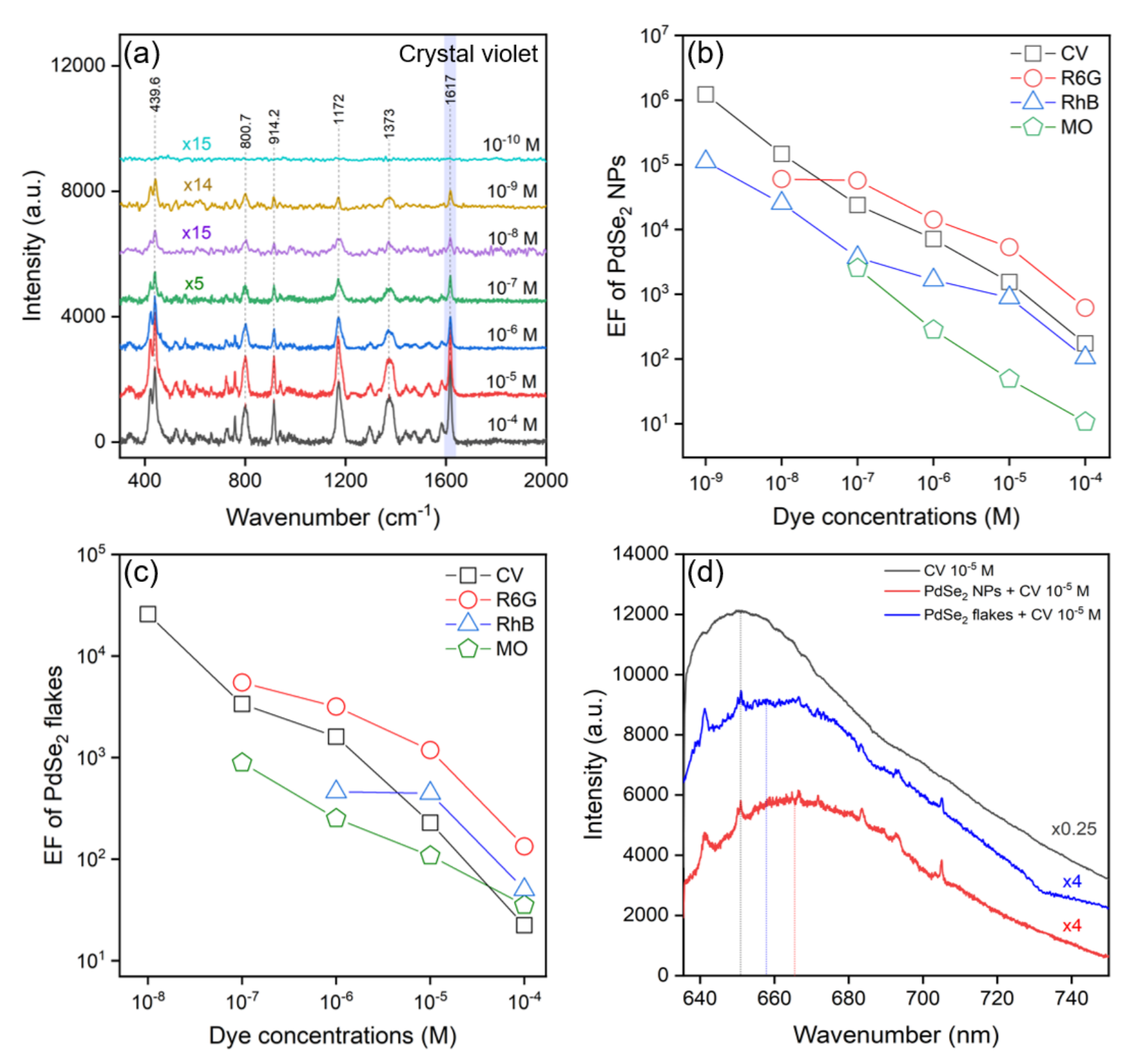}
\caption{(a) SERS spectra of CV acquired using PdSe\textsubscript{2-x} NP-based substrates; b),c) Enhancement factors (EF) calculated for NP-based PdSe\textsubscript{2-x} (b) and flake-based PdSe\textsubscript{2} (c) substrates across all tested dyes; (d) Measured photoluminescence spectra of $10^{-5}$ M CV on glass substrate, SERS spectrum of NP-based PdSe\textsubscript{2-x}/$10^{-5}$ CV and flake-based PdSe\textsubscript{2}/$10^{-5}$ CV.}
\label{Article4}
\end{figure}

The most representative SERS spectra, which were measured with crystal violet on a PdSe\textsubscript{2-x} NP-based substrate, are shown in Figure \ref{Article4}a. As anticipated, the SERS intensity decreased progressively with lower dye concentrations. At concentrations as low as 1 × $10^{-9}$ M for NP-based substrates, dye peaks remained detectable within the 400 to 1700 cm$^{-1}$ range, although with reduced intensity (see Figure \ref{Article4}a). This indicates that the limit of detection (LOD) of the PdSe\textsubscript{2-x} NP-based substrates was $10^{-9}$ M. Similar LOD values were also achieved for Rh6G, RhB and MO.

Recent studies \cite{jena2023evidence,lei2022enhanced} have confirmed the high efficiency of SERS sensors utilizing PdSe\textsubscript{2} structures, reporting LOD values ranging from $10^{-6}$ to $10^{-9}$ M. These findings underscore the potential of PdSe\textsubscript{2}-based substrates for sensitive SERS applications.

To evaluate the performance of SERS of both substrate types, we determined the enhancement factor (EF) defined as EF=I\textsubscript{SERS}/I\textsubscript{RS}$\times$C\textsubscript{RS}/C\textsubscript{SERS}, where I\textsubscript{SERS} and I\textsubscript{RS} are the signal intensities from SERS and conventional Raman measurements, respectively, and C\textsubscript{SERS} and C\textsubscript{RS} represent the analyte concentrations corresponding to these intensity measurements. The calculated enhancement factors for each substrate are presented in Figure \ref{Article4}b, c and summarized in Table \ref{Table1}.

\begin{table}[H]
\resizebox{\textwidth}{!}{%
\centering
\begin{tabular}{clcccccc} 
\hline\hline

\begin{tabular}[c]{@{}c@{}}Dye\\concentrations\end{tabular} &  & Crystal Violet & Rhodamine B & Rhodamine 6G & \multicolumn{2}{c}{Methyl Orange} & \\
               &  & \textbf{1617 cm$^{-1}$}        & \textbf{1649 cm$^{-1}$}        & \textbf{607 cm$^{-1}$}          & \textbf{1113 cm$^{-1}$}           & \textbf{1139 cm$^{-1}$} &   \\ 
\cline{1-1}\cline{3-7}
10$^{-4}$ M         &  & 170               & 100         & 620           & 10             & 10          &   \\ 
\cdashline{1-1}\cdashline{3-7}
10$^{-5}$ M         &  & 1550              & 890        & 5330          & 40            & 50           &   \\ 
\cdashline{1-1}\cdashline{3-7}
10$^{-6}$ M         &  & 7130             & 1660        & 14200          & 300            & 280         &   \\ 
\cdashline{1-1}\cdashline{3-7}
10$^{-7}$ M         &  & 23930             & 3680        & 57210        &2150            & 2520          &   \\
\cdashline{1-1}\cdashline{3-7}
10$^{-8}$ M         &  & 147290            & 25410       & 60110         & ND              & ND          &   \\
\cdashline{1-1}\cdashline{3-7}
10$^{-9}$ M         &  & 1226420          & 111570         & ND          & ND              & ND          &   \\
\hline\hline
\end{tabular}}
\caption{SERS enhancement factors for PdSe\textsubscript{2-x} nanoparticles at varying dye concentrations. "ND" (Not detectable) symbols indicate values below the detection limit.}
\label{Table1}
\end{table}

 The results demonstrate that PdSe\textsubscript{2-x} NPs can serve as effective SERS substrates, enabling the detection of dyes at low concentrations ($10^{-9}$ M) with EF reaching as high as $10^{6}$. In contrast, crystalline PdSe\textsubscript{2} flakes are significantly less effective, requiring higher dye concentrations ($10^{-7}-10^{-8}$ M) for detection with EF $\sim 10^{4}-10^{5}$, depending on the specific dye. These results correlate with previous reports \cite{lei2022enhanced, jena2023evidence, jena2024salt}.

The origin of the exceptionally high enhancement factor of $10^{6}$ observed for CV on PdSe\textsubscript{2-x} nanoparticle substrates is a combination of synergistic effects that go beyond the traditional chemical mechanism (CM) of SERS enhancement. CM typically provides EFs in the range of $10^{2}$–$10^{4}$ \cite{persson2006chemical}, so the significantly higher enhancement seen in this system arises from several interconnected processes. One key factor is the resonant charge transfer between CV molecules and PdSe\textsubscript{2-x} nanoparticles, facilitated by the alignment of energy levels. When illuminated, electrons from the dye's molecular orbitals can efficiently transfer to the conduction band of PdSe\textsubscript{2-x}, creating a charge-transfer complex that dramatically increases Raman scattering cross-sections \cite{lei2022enhanced,jena2023evidence, kang2023amorphous}.

The unique electronic structure of PdSe\textsubscript{2} plays a crucial role in this enhancement. Unlike conventional plasmonic metals, PdSe\textsubscript{2} possesses a tunable optical bandgap that can be engineered to match the electronic transitions of target molecules such as CV. This energy matching creates optimal conditions for resonant enhancement, where both the incident laser light and the Raman scattered photons interact strongly with the charge-transfer states. Furthermore, the two-dimensional nature of PdSe\textsubscript{2-x} NPs provides a large surface area with numerous active sites for molecular adsorption, which further amplifies the signal.

Defects and nanostructural features on the PdSe\textsubscript{2-x} substrate also contribute significantly to the observed enhancement \cite{jena2023evidence, majumdar20242d, zhao2025metallic}. Selenium vacancies and edge sites in the nanoparticles create localized electronic states that act as efficient charge traps \cite{xu2022selenium}. These defect sites not only facilitate charge transfer processes but may also generate localized electromagnetic hotspots through dielectric confinement effects. The nanoscale roughness and particle aggregation create electromagnetic field enhancements similar to those observed in traditional plasmonic systems, despite the absence of conventional plasmon resonance in PdSe\textsubscript{2-x} \cite{jena2023evidence}.

The combination of these effects -- resonant charge transfer, defect-mediated enhancement, and nanoscale field concentration -- creates a multiplicative effect that explains the unusually high EF values \cite{berestennikov2025molecular}. This is particularly evident in the fluorescence quenching shown in Fig.~\ref{Article4}d, where the suppression of radiative decay may be attributed to efficient charge transfer. The result is a system where Raman signals are dramatically enhanced, while fluorescence is minimized, creating ideal conditions for sensitive SERS detection \cite{jena2023evidence}. 

The lower enhancement factor observed for PdSe\textsubscript{2} flakes ($10^{5}$) (see Fig.~\ref{Article4}c) compared to nanoparticles ($10^{6}$) can be attributed to several key structural and morphological differences. Unlike the amorphous nanoparticles, the flakes possess a well-ordered crystalline structure with a much lower number of defects and vacancies. While this crystalline perfection is advantageous for many electronic applications, it actually hinders SERS performance by limiting the availability of active sites for dye adsorption and charge transfer processes \cite{kang2023amorphous}. 

The nanoparticles' amorphous nature, with their abundant defects and non-stoichiometric composition (Pd:Se $\sim$ 1:1), creates numerous favorable sites for molecular interaction and charge transfer enhancement \cite{kang2023amorphous}. In contrast, stoichiometric flakes (Pd:Se $\sim$ 1:2) possess fewer broken Se bonds on the surface compared to the nanoparticles. Missing Se atoms in the nanoparticle structure create vacancies that serve as ideal anchoring points for dye molecules such as crystal violet. The more perfect surface of the flakes, while structurally superior, lacks these natural adsorption sites, making it more difficult for dye molecules to attach and participate in the charge transfer processes that amplify SERS signals. This difference in surface chemistry directly impacts the efficiency of the chemical enhancement mechanism.

The morphological differences between the flakes and the nanoparticles also play a significant role. The two-dimensional, planar structure of the flakes offers a much smaller specific surface area compared to that of the three-dimensional nanoparticles. This reduced surface area means that fewer sites are available for dye adsorption, further limiting the overall enhancement. 

Fluorescence quenching, which is so prominent in the nanoparticle samples, is also less pronounced in the flake structures (see Fig.~\ref{Article4}d). The reduced defect density and different surface chemistry of the flakes result in less effective competition between charge transfer pathways and radiative decay processes. This explains why the fluorescence background is more noticeable in flake-based SERS measurements compared to that of nanoparticle samples, where charge transfer dominates the relaxation pathways. 

The variation in SERS signals observed for different dyes (Rh6G, RhB, CV, and MO) on PdSe\textsubscript{2-x} NP-based substrates stems from a combination of molecular adsorption, charge transfer efficiency, and resonance effects. Planar aromatic dyes like Rh6G, RhB, and CV exhibit stronger SERS signals due to their ability to $\pi$-stack efficiently with the PdSe\textsubscript{2-x} surface, facilitated by Se vacancies that act as preferential adsorption sites \cite{canamares2008dft}. Crystal violet shows particularly strong enhancement owing to its triple aromatic ring system and N-dimethyl groups that enable multi-point binding to Pd sites, while Rh6G and RhB benefit from favorable energy alignment of the lowest unoccupied molecular orbital with the substrate's conduction band, enabling resonant charge transfer under typical excitation wavelengths. In contrast, methyl orange's non-planar structure and sulfonate group hinder effective adsorption and charge transfer, resulting in weaker signals \cite{laguna2007azoic}. The amorphous nature of the PdSe\textsubscript{2-x} nanoparticles further enhances these effects through defect-mediated mechanisms: Se vacancies create mid-gap states that selectively boost dye-specific charge transfer when electronic transitions align with the laser excitation. Additionally, dye aggregation plays a role; Rh6G's tendency to form J-aggregates on nanoparticle surfaces creates localized electromagnetic hotspots even without plasmonic effects. 

These differences highlight how molecular structure dictates interaction strength with the substrate, where planar, electron-rich dyes with compatible energy levels achieve optimal enhancement through combined chemical and defect-assisted mechanisms. The substrate's selectivity for specific dye architectures suggests potential for tailored sensing applications, where the SERS response could be engineered by controlling the nanoparticle defect density and surface chemistry to target particular molecular features. 

These findings suggest that careful engineering of two-dimensional material substrates could open new possibilities for ultra-sensitive molecular detection beyond the limitations of conventional SERS platforms \cite{he2012surface,sil2013chemically,watanabe2005dft}.

\subsection{Photocatalytic activity of palladium selenide nanoparticles and flakes}

\begin{figure}[H]
\centering\includegraphics[width =1\textwidth]{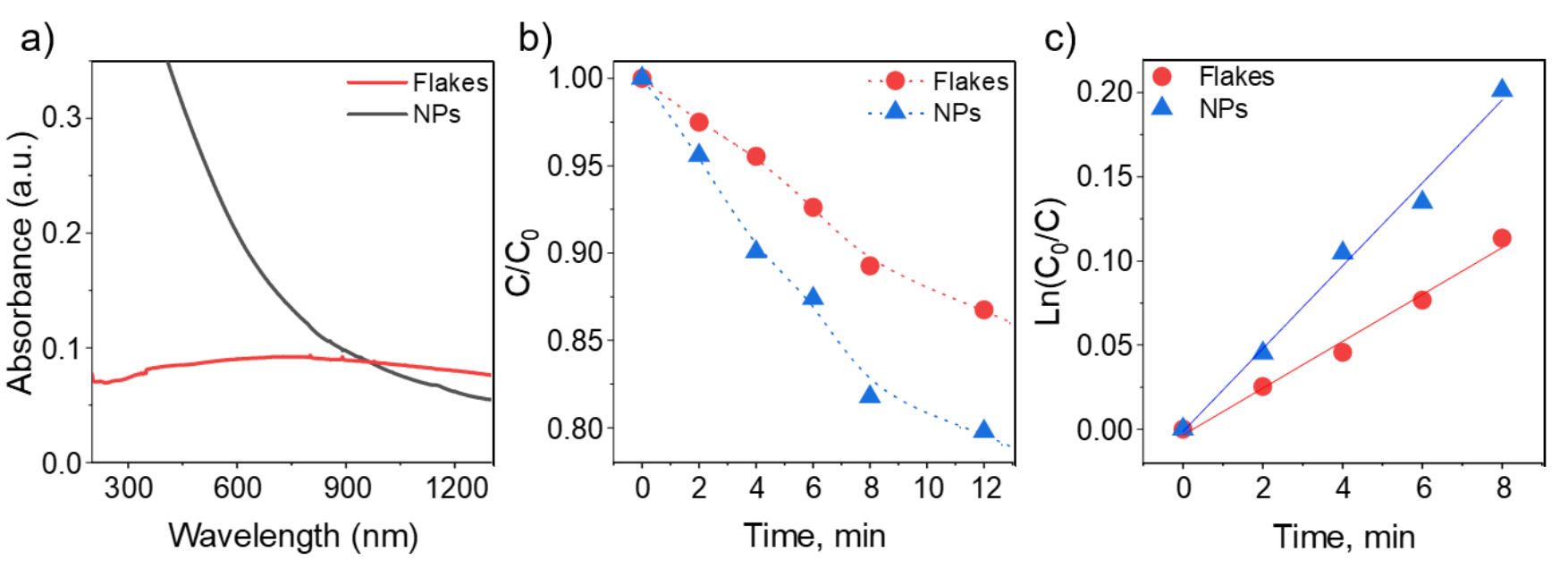}
\caption{(a) UV–visible absorption spectra of PdSe\textsubscript{2-x} nanoparticles and PdSe\textsubscript{2} flakes; (b) The dependence of the normalized MB concentration (C\textsubscript{0}/C\textsubscript{t}) on irradiation time for both catalysts; (c) Fitting of ln(C\textsubscript{t}/C\textsubscript{0}) versus time.}
\label{Article5}
\end{figure}

Figure \ref{Article5} provides a comprehensive analysis of the photocatalytic activity of amorphous PdSe\textsubscript{2-x} nanoparticles and PdSe\textsubscript{2} flakes in the degradation of methylene blue (MB) under simulated solar irradiation (100 mW/cm$^{2}$, AM 1.5G). 

Figure \ref{Article5}a displays the UV-Vis absorption spectra for both materials. The amorphous PdSe\textsubscript{2-x} nanoparticles, at a concentration of 0.022 mg/mL, exhibit a broad absorption band extending into the visible region, attributed to defect states induced by selenium vacancies. These defects enhance solar energy capture, which is crucial for photocatalysis under visible light. In contrast, PdSe\textsubscript{2} flakes, at a much higher concentration of 0.66 mg/mL, show a sharper absorption edge, typical of their crystalline structure and bandgap absorption, with minimal defect contribution.

Figure \ref{Article5}b illustrates the dependence of the normalized MB concentration (C\textsubscript{t}/C\textsubscript{0}) on irradiation time for both catalysts. Despite their lower concentration, PdSe\textsubscript{2-x} nanoparticles achieve fast MB degradation within 12 minutes, while the flakes, with a 30-fold higher concentration, require more time to reach a comparable level of degradation. The rapid degradation by the nanoparticles underscores their superior efficiency per unit mass, highlighting the impact of morphology and defect states.

Figure \ref{Article5}c confirms the pseudo-first-order kinetics of MB degradation through linear dependencies of ln(C\textsubscript{0}/C\textsubscript{t}) on time. The calculated rate constants are 0.0244 ± 0.0007 min$^{-1}$ for the nanoparticles and 0.0138 ± 0.0005 min$^{-1}$ for the flakes. These values reveal the nanoparticles’ enhanced photocatalytic efficiency, even at a significantly lower concentration. Normalizing the rate constants by concentration further emphasizes this disparity: the nanoparticles exhibit two orders of magnitude higher specific activity (k/concentration) of approximately 1.11 min$^{-1}$·mL/mg, compared to 0.021 min$^{-1}$·mL/mg for the flakes—a 50-fold higher efficiency per unit mass. This can be attributed to the nanoparticles’ increased surface area, improved charge transfer dynamics due to defects, and the stronger light absorption.
Overall, the data highlight the critical role of morphology, defect structure, and concentration in determining the photocatalytic performance of PdSe\textsubscript{2-x}. Amorphous nanoparticles not only demonstrate broader light absorption but also achieve significantly higher degradation rates under visible light, even at a fraction of the concentration of the flakes. These findings suggest that optimizing material morphology and defect engineering can substantially enhance the photocatalytic efficiency, opening prospects for the practical application of PdSe\textsubscript{2-x} in sustainable technologies, such as water purification.

\section{Conclusion}

In conclusion, we introduced PLAL as a material processing tool for fabrication of stable amorphous nanomaterials (PdSe\textsubscript{2-x}) from the crystalline PdSe\textsubscript{2} target. The synthesized nanoparticles achieve a record-high EF$>10^6$ for CV and $>10^5$ for RhB, what exceeds the crystalline flakes by more than an order of magnitude. The measured photocatalytic activity of PdSe\textsubscript{2-x} NPs demonstrates a 50-fold higher efficiency per unit mass, which is attributed to the nanoparticles’ increased surface area, improved charge transfer dynamics due to defects, and stronger light absorption. Interestingly, the nanoparticles inherit strong optical absorption of the crystalline PdSe\textsubscript{2} precursor. Their high photothermal conversion up to $\eta=83\%$ puts them on par with recognized highly efficient photoheating agents made of TiN and Ti\textsubscript{3}C\textsubscript{2}T\textsubscript{x}. Our findings establish amorphous PdSe2-x NPs as a promising multipurpose platform for plasmon-free sensing, photocatalysis and photoheating applications.

\section{Experimental Section}

\subsection{Nanoparticles synthesis via femtosecond laser ablation}
\label{label_materials_and_methods}

PdSe\textsubscript{2-x} nanoparticles were synthesized via pulsed laser ablation in liquid (PLAL) using a femtosecond laser. A bulk PdSe\textsubscript{2} crystal served as the ablation target. In the PLAL process, the target material absorbs laser pulse energy through nonlinear optical processes, leading to plasma plume formation and subsequent confinement within a cavitation bubble. The subsequent bubble growth and collapse dynamics strongly depend on the surrounding liquid properties (i.e., liquid density, heat capacity and conductivity, boiling point), defines the plasma condensation rate and, consequently, the resulting NPs content and crystallinity.

The ytterbium-doped potassium gadolinium tungstate (Yb:KGW) laser system (TETA-10, Avesta, Moscow, Russia) was used as a source of femtosecond pulses (wavelength 1030 nm, pulse duration ~270 fs, repetition rate 1 kHz); the pulse energy for PLAL was set at \SI{100}{\micro\joule}. A synthetic bulk PdSe\textsubscript{2} crystal (dimensions: 3 × 4 × 4 mm, SixCarbon Technology Co., Limited, ShenZhen) was used as the ablation target and the source of the crystalline flakes. The ultrasonic-assisted liquid phase exfoliation method \cite{chavalekvirat2024liquid} was employed for PdSe\textsubscript{2} flakes production. Under this approach, the probe tip sonicator (CL-18 benchtop ultrasonic processor, Qsonica L.L.C., Newtown, USA) delivers ultrasonic energy (100 W for 4 hours) directly to the bulk crystal fragment, weighted in advance to get a mass concentration of 2 mg/ml.

For the pulsed laser ablation process, a bulk PdSe\textsubscript{2} crystal was placed at the bottom of a glass cuvette filled with 2 ml of deionized water as the ablation medium. The laser beam was focused on the target surface using a 100-mm focal lens, creating a spot diameter of $\sim$ 50 \textmu m. To ensure uniform ablation across the target, the cuvette was translated in a 2 $\times$ 2 mm area using two motorized stages (scan speed: 5 mm/s), thereby exposing the entire crystal surface to the laser. The pulse energy was fixed at \SI{100}{\micro\joule}, corresponding to a fluence of $\sim$ 5 J/cm\textsuperscript{2} at the focal spot. A series of PLAL experiments were conducted by varying the distance between the focal plane and the crystal surface. In each trial, the ablation was limited to 15 minutes. Within minutes of irradiation, the solution developed a visible tint, confirming the successful generation of PdSe\textsubscript{2-x} nanoparticles.

After laser ablation, colloid remained stable and did not show visible changes in color for at least a month. The samples were stored in closed 2 ml Eppendorf microcentrifuge tubes at 6\textcelsius{} to prevent contamination or evaporation. 

To narrow the size distribution of the synthesized PdSe\textsubscript{2-x} NPs, we employed differential centrifugation. The nanoparticle colloid was centrifuged (Eppendorf centrifuge 5424 R, Hamburg, Germany) at progressively increasing speeds from 500 to 4000 rpm, enabling separation by size.

\subsection{Sample characterization}

\subsubsection{Microscopy and energy-dispersive X-ray spectroscopy}

Nanoparticle colloids were characterized by the high-resolution TEM system (JEOL JEM 2100, Japan) with an acceleration voltage of 200 kV. 2-\textmu l drops of sample colloids were drop-casted on carbon-coated TEM copper grids and dried under ambient conditions.

A scanning SEM system (MAIA 3; Tescan) with an integrated EDX detector (X-act; Oxford Instruments) was used to examine a bulk PdSe\textsubscript{2} crystal surface before and after ablation.

\subsubsection{Photoluminescence spectroscopy}

Photoluminescence spectroscopy was performed on colloidal suspensions of PdSe\textsubscript{2} flakes and PdSe\textsubscript{2-x} nanoparticles using a BioTek Synergy H4 Hybrid Multi-Mode Microplate Reader (BioTek Instruments, Inc., Vermont, USA). Spectra were acquired at room temperature under controlled excitation conditions.

\subsubsection{X-ray photoelectron spectroscopy}

We analyzed the surface composition and optical band gap of both bulk PdSe\textsubscript{2} crystals and synthesized PdSe\textsubscript{2-x} nanoparticles using X-ray photoelectron spectroscopy (XPS). Measurements were performed with a Theta Probe instrument (Thermo Scientific) under high-vacuum conditions, employing a monochromatic Al-K$\alpha$ X-ray source (1486.6 eV). All spectra were recorded in fixed analyzer transmission mode with a pass energy of 50 eV, and the spectrometer energy scale was meticulously calibrated using the Au 4f\textsubscript{7/2} line at 84.0 eV.
The XPS spectra for Pd\textsubscript{3d}, O\textsubscript{1s} and Se\textsubscript{3d} were measured along with the spectra of the valence bands. For nanoparticle characterization, samples were prepared by depositing 1 mL of the NP solution onto silicon substrates, followed by drying under ambient conditions.

\subsubsection{Photothermal studies}

Photoheating experiments were performed with a tunable titanium-sapphire laser source at a NIR-I wavelength of 830 nm. The temperature dynamics was monitored in real time using a calibrated HIKMICRO M10 thermal imaging camera. Colloidal extinction was measured using transmission through a cuvette with deionized water as a baseline.

\subsubsection{Raman spectroscopy and SERS}
 
For surface-enhanced Raman spectroscopy (SERS) measurements, 2 \textmu l of sample colloids were drop casted onto an aluminum surface and spin-coated to form thin films. The prepared SERS substrates were then coated with 2 \textmu l of dye solutions at various concentrations ranging from
$10^{-4}-10^{-10}$ M and allowed to dry under ambient conditions for 1 hour. Raman spectroscopy was carried out using a Horiba LabRAM HR Evolution system equipped with excitation wavelengths of 532 and 633 nm, a diffraction grating of 600 lines/mm, and a 100$\times$ microscope objective with a numerical aperture (NA) of 0.9. The measurements demonstrated good spectral reproducibility.

\subsubsection{Photocatalytic Studies}

\textit{Preparation of photocatalytic Suspensions}

Amorphous PdSe\textsubscript{2-x} nanoparticles were synthesized and dispersed in deionized water to form a stable colloidal suspension. The suspension was adjusted to a concentration of 0.022 mg/mL and combined with a 20 mg/L methylene blue solution (Sigma-Aldrich, purity 98\%) in quartz tubes. For comparative studies, PdSe\textsubscript{2} flakes were prepared at a concentration of 0.66 mg/mL in identical MB solutions. All chemicals were used as received without further purification.

\textit{Photocatalytic experiments}

To ensure adsorption–desorption equilibrium, the suspensions were stirred continuously at 500 rpm in the dark for 30 minutes at 30 °C using a magnetic stirrer. Photocatalytic degradation was initiated by exposing the suspensions to simulated solar irradiation generated by a solar simulator (Numi Technology ESS-500-a, 100 mW/cm$^{2}$, AM 1.5G spectrum). The light intensity was calibrated using a silicon reference cell. The reaction mixtures were constantly stirred to ensure uniform irradiation and mass transfer.

\textit{Sample analysis}

At designated time intervals (0, 2, 4, 6, 8, and 12 minutes), 1 mL aliquots were taken from the reaction mixture and immediately centrifuged at 15000 rpm for 10 minutes (Eppendorf 5424R centrifuge) to separate the nanoparticles. The supernatant was carefully decanted and analyzed using a UV-Vis spectrophotometer (Agilent Cary 5000) to measure the absorbance of MB at its characteristic peak of 665 nm. The photodegradation efficiency was calculated as the ratio of MB concentration at time t (C\textsubscript{t}) to the initial concentration at adsorption equilibrium (C\textsubscript{0}), based on the Lambert–Beer law.

\textit{Kinetic analysis}

The kinetics of MB photodegradation were modeled using a pseudo-first-order rate equation: ln(C\textsubscript{0}/C\textsubscript{t}) = kt, where k is the apparent rate constant (min$^{-1}$) and t is the irradiation time (min). Experimental data were fitted to determine the rate constants and assess the photocatalytic performance of the PdSe\textsubscript{2-x} nanoparticles and flakes. All experiments were performed in triplicate and the results were reported as mean values with standard deviations.

\section*{Acknowledgments}
The authors acknowledge Alexander Ushkov for the graphical illustration of the PLAL setup.

\section*{Conflicts of Interest}
The authors declare no conflict of interest.

\section*{Author Contributions}
A.A. conceived the idea, supervised the project and edited the manuscript. A.U. designed and performed the experiments and wrote the manuscript. D.D. performed the experiments. N.B. conducted the characterizations, performed the experiments and wrote the manuscript. I.M. designed and performed the experiments and wrote the manuscript. A.S. conducted the TEM/SAED characterizations. D.T. performed the experiments. G.E., S.V.B. and R.I.R. conducted characterizations. I.K. performed DFT simulations. A.A.P., A.C. and S.N. provided experimental resources. A.D.B., G.I.T., A.V.K. and V.V. supervised the project. A.A.V. supervised the project and edited the manuscript. All authors discussed the results. All authors have read and agreed to the published version of the manuscript.

\section*{Data Availability Statement}
The data that support the findings of this study are available from the corresponding author upon reasonable request.

\section*{Supporting Information}

\section*{Abbreviations}
The following abbreviations are used in this manuscript:\\

\noindent 
\begin{tabular}{@{}ll}
vdW & Van der Waals\\
DI water & Deionized water\\
TMDC & Transition Metal Dichalcogenide\\
NP & Nanoparticle\\
PLAL & Pulsed Laser Ablation in Liquid\\
DFT & Density functional theory\\
SEM & Scanning Electron Microscope\\
TEM & Transmission Electron Microscope\\
SAED & Selected Area Electron Diffraction\\
EDX & Energy-dispersive X-ray spectroscopy\\
PL & Photoluminescence\\
XPS & X-Ray Photoelectron Spectroscopy\\
SERS & Surface-Enhanced Raman Scattering\\
LOD & Limit of Detection\\
EF & Enhancement factor\\
Rh6G & Rhodamine 6G\\
RhB & Rhodamine B\\
MO & Methyl Orange\\
CV & Crystal Violet\\
MB & Methylene Blue\\
NA & Numerical Aperture\\
EPR & Enhanced Permeability and Retention\\
PCE & Photothermal Conversion Efficiency
\end{tabular}

\bibliographystyle{elsarticle-num}
\bibliography{references.bib}

\end{document}